\documentclass[prl,twoside,tightenlines,superscriptaddress,twocolumn,aps, 10pt]{revtex4}
\usepackage{bbm,amssymb,amsfonts,amsthm,amsmath,graphicx}

\newcommand{\cH}{\mathcal{H}}

\newcommand{\ket}[1]{\left| #1\right\rangle}      
\newcommand{\bra}[1]{\left\langle #1\right|}      
\newcommand{\kets}[1]{| #1 \rangle}                 
\newcommand{\ii}{\mathbb{I}}
\newcommand{\ep}{\epsilon}

\begin{document}
	\title{Universal 2-local Hamiltonian Quantum Computing}
		\author{Daniel Nagaj}
		\email{daniel.nagaj@savba.sk}
		\affiliation{Research Center for Quantum Information, Institute of Physics, Slovak
		Academy of Sciences, D\'ubravsk\'a cesta 9, 84215 Bratislava, Slovakia}

\begin{abstract}
We present a Hamiltonian quantum computation scheme universal for quantum computation (BQP). 
Our Hamiltonian is a sum of a polynomial number (in the number of gates $L$ in the quantum circuit) of time-independent, constant-norm, 2-local qubit-qubit interaction terms. Furthermore, each qubit in the system interacts only with a constant number of other qubits. The computer runs in three steps -- starts in a simple initial product-state, evolves it for time of order $L^2$ (up to logarithmic factors)
and wraps up with a two-qubit measurement. Our model differs from the previous universal 2-local Hamiltonian constructions in that it does not use perturbation gadgets, does not need large energy penalties in the Hamiltonian and does not need to run slowly to ensure adiabatic evolution.
\end{abstract}

\maketitle
	


Part of today's effort at achieving a realization of a quantum computer 
is turning away from the traditional quantum circuit model with sequential application of unitary gates \cite{NCbook}. Instead, measurement-based \cite{model:measurement} computation, cluster state quantum computation \cite{model:cluster}, topological quantum computation \cite{model:anyons} computation by quantum walks \cite{model:walk}, adiabatic quantum computation (AQC) \cite{model:AQC} 
and the usage of adiabatic gate teleportation \cite{AdiabaticGateTeleport} are some of the recently explored alternatives. The benefit of AQC is that one does not require precise and fast control over logical operations and measurements. Instead, it relies on slow change of the Hamiltonian, keeping the system in its ground state. The proofs of universality of AQC \cite{universalAQC}\cite{KempeKitaevRegev}\cite{LidarMizel} rely on techniques from Kitaev's QMA-complete ''local Hamiltonian'' problem \cite{KitaevBook}. The ground state of the final Hamiltonian contains the result of the intended computation, and we prepare it adiabatically.

Universal quantum computation can also be implemented efficiently with a system with a time independent interactions. Such a Hamiltonian quantum computer (HQC, or ergodic q. c. \cite{JW:05}) runs in three stages. First, 
it starts in a simple initial computational-basis product state. Second, the system undergoes Schr\"{o}dinger time evolution for some time. Finally, we measure a few of the qubits in the computational basis to obtain the answer to the computation. However, so far the universal systems involved at least 3-local (long-range) interactions \cite{Bravyi}\cite{New3local}, or 2-local nearest neighbor interactions (on a chain) of high-dimensional particles (qudits) 
\cite{AGKIline}\cite{other1D}\cite{ERtriangle}.
Lloyd \cite{Lloyd} has shown that the HQC model can be universal, not relying on changing the system adiabatically. As long as the time evolution runs (as a quantum walk) in an invariant computational subspace,
the relevant excited states of the Hamiltonian involved also contain the result of the computation. 

Until now, the only universal quantum computation with a 2-local Hamiltonian was an AQC model based 
on \cite{KKR}\cite{OliveiraTerhal} (with restricted terms in \cite{BiamonteLove}), with a gap over the ground state lower bounded by a high-degree inverse polynomial in the circuit size $L$. The runtime of the model is thus necessary a high-degree polynomial in $L$. What is worse, the strengths of the interaction terms in the Hamiltonian are high-degree polynomials in $L$.

We present a new universal 2-local Hamiltonian quantum computer construction in the HQC model with a polynomial (in $L$) number of constant-norm interaction terms. We achieve 2-locality by a combining the railroad switch \cite{NagajSwitch} and entangled-clock \cite{New3local} ideas, 
ensuring the evolution of a simple initial product state 
does not leave the computational subspace. 


\section{Evolution within a good subspace}

Consider a quantum circuit $U=U_L U_{L-1} \dots U_2 U_1$ with $L$ at most 2-local unitary gates.
We would like to obtain the result of $U$ acting on some $n$-qubit initial state $\ket{\varphi_0}$,
i.e. to measure the first (output) qubit of the state $U\ket{\varphi_0}$. 
Instead of working with just the $n$ ``work'' qubits of $\ket{\varphi_0}$, we utilize a quantum system with two registers $\cH_{w}\otimes \cH_{c}$, work and clock. The work register holds the work qubits and the clock register contains pointer states corresponding to the progress of the computation.
Consider now the ``line'' of states
\begin{eqnarray}
	\ket{\psi_t} = \left( U_t U_{t-1} \dots U_2 U_1 \ket{\varphi_0} \right) \otimes \ket{t}_c
	\label{goodbasis}
\end{eqnarray}
for $t=0,\dots,L$.
These states encode the progress of the quantum circuit $U$ acting on the initial state $\ket{\varphi_0}$,
and the state $\ket{\psi_L}$ contains the result of the quantum circuit acting on $\ket{\varphi_0}$.
Denote the span of these states
$
	\cH_{\mathrm{cmp}}^{\varphi_0} = \textrm{span} \left\{ \ket{\psi_t} \right\}
	\label{goodspace}
$ and call it the {\em computational subspace}.
Our approach is to use a system whose Hamiltonian does not induce transitions out of $\cH_{\mathrm{cmp}}^{\varphi_0}$. This makes the dynamics of this model and its required running time simple to analyze.

\begin{figure}
	\begin{center}
	\includegraphics[width=1.5in]{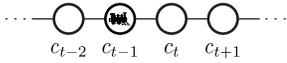} 
	\end{center}
	\caption{The {\em pulse clock} -- a line of $L+1$ qubits $c_t$
	with a single active site (train). 
	The states $\ket{t}_c$ are encoded as $\ket{0\cdots 010\cdots 0}$,
	with qubit $c_t$ in the state $\ket{1}$.
	When the train moves from $c_{t-1}$ to $c_{t}$, gate $U_t$ is applied to two work qubits.
	\label{figurepulse}}
\end{figure}
We start with Feynman's Hamiltonian \cite{Feynman85}:
\begin{eqnarray}
	H_{F} =  \sum_{t=1}^{L} \underbrace{\left(
		   U_t \otimes \ket{t}\bra{t-1}_{c}
		+  U_t^\dagger \otimes \ket{t-1}\bra{t}_{c} \right)}_{H_F^{(t)}},
	\label{HFeynman}
\end{eqnarray}
Observe that the computational subspace is invariant
under $H_F$. The restriction of $H_{F}$ to $\cH_{\mathrm{cmp}}^{\varphi_0}$ is
\begin{eqnarray}
	H_{F}\big|_{\cH_{\mathrm{cmp}}^{\varphi_0}} = 
	 \sum_{t=1}^{L} \left(
	 	\ket{\psi_t}\bra{\psi_{t-1}} + \ket{\psi_{t-1}}\bra{\psi_{t}}
	 \right),
	 \label{Hline}
\end{eqnarray}
a quantum walk on the ``line'' of states $\ket{\psi_t}$. 
We now use a {\em pulse clock} encoding 
$\ket{t}_c = \ket{0\cdots0_{c_{t-1}}1_{c_t}0_{c_{t+1}}\cdots0}$
of the clock register states, using $L+1$ qubits
(see Figure \ref{figurepulse}).
We can make the terms in the Hamiltonian \eqref{HFeynman} at most 4-local,
as the gates $U_t$ are at most 2-local and we can use 2-local operators for the clock-register transitions
\begin{eqnarray}
	\ket{t}\bra{t-1}_c &=& \ii \otimes \ket{01}\bra{10}_{c_{t-1},c_{t}} \otimes \ii.
\end{eqnarray}
Writing it like this, we obtain a 4-local Hamiltonian different from $H_F$. However, its restriction to 
$\cH_{\mathrm{cmp}}^{\varphi_0}$ is again \eqref{Hline},
generating the desired quantum walk on a line starting with the initial state $\ket{\psi_0} = \ket{\varphi_0}\ket{0}_c$. 
The last step of the HQC model is then to measure the clock register and the output work qubit.
If we find the clock register in the state $\ket{L}_c$, we obtain the answer to the computation.
We boost the probability to actually measure $\ket{L}_c$ (or in our case, $\ket{1}_{c_L}$) 
by adding many identity gates at the end of the circuit $U$. This means 
measuring 
$\ket{1}_{c_{t>L}}$ is enough to ensure the output work qubit holds the output of $U$,
and is thus enough to ensure BQP universality.

\begin{figure}
	\begin{center}
	\includegraphics[width=2.0in]{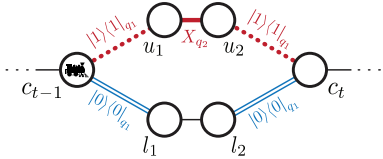} 
	\end{center}
	\caption{The 3-local {\em railroad switch} gadget for the application of a CNOT gate between work qubits $q_1$ and $q_2$. 
The state of the control (train master) qubit $q_1$ in the work register decides whether the train moves to the upper/lower track from $c_{t-1}$ (and backwards from $c_{t}$). On the upper track, we flip the target work qubit $q_2$ when the train moves from $u_1$ to $u_2$. 
	\label{figureswitch}}
\end{figure}

\section{The railroad switch}

We now modify the clock register, introducing a {\em railroad switch} gadget \cite{NagajSwitch,Feynman85}. This will give us a 3-local Hamiltonian equal to
\eqref{Hline} when restricted to the computational subspace.
The pulse clock (see Figure \ref{figurepulse}) can be viewed as a train running on a single track. When it goes between stations $c_{t-1}$ and $c_{t}$, the transition in $H_F$ ensures the gate $U_t$ is applied to the corresponding work qubits in the work register.
The railroad switch (see Figure \ref{figureswitch}) introduces four extra stations between $c_{t-1}$ and $c_{t}$, allowing the train to run on two tracks.
The train is allowed to move to the upper or lower track depending on the state of a ``train master'' -- one of the work qubits $q_1$. Furthermore, the target work qubit $q_2$ is flipped on the upper track as the train moves from $u_1$ to $u_2$. The computational paths running on the upper and lower tracks interfere at station $c_{t}$. This gadget facilitates the application of a CNOT gate on the work qubits $q_1$ and $q_2$.

Let us look at the dynamics coming from this Hamiltonian. Denote the extra clock states (train positions) between $\ket{t-1}_c$ and $\ket{t}_c$
as $\ket{u_1}_c, \ket{u_2}_c, \ket{l_1}_c, \ket{l_2}_c$. We replace $H_F^{(t)}$ in \eqref{HFeynman} by
\begin{eqnarray}
	H_{\mathrm{sw}}^{(t)} &=& 
		\ket{0}\bra{0}_{q_1} \otimes \left( \ket{l_1}\bra{t-1}_c + \ket{t}\bra{l_2}_c \right) \nonumber\\
	&+& \ket{1}\bra{1}_{q_1} \otimes \left( \ket{u_1}\bra{t-1}_c + \ket{t}\bra{u_2}_c \right) \label{Hswitch}\\
	&+& X_{q_2} \otimes \ket{u_2}\bra{u_1}_c + \ii\otimes\ket{l_2}\bra{l_1}_c + c.c.  \nonumber
\end{eqnarray}
We make all of the terms 3-local by writing each clock transition 
such as $\ket{l_1}\bra{t-1}_c$ only 2-locally as 
$\ket{01}\bra{10}_{l_{1},c_{t-1}}$.
Each term in $H_{switch}^{(t)}$ then acts nontrivially on 
at most one work qubit ($q_1$ or $q_2$), and two clock qubits.
We now augment the 
``line'' of states \eqref{goodbasis}, taking into account the intermediary states we introduced.
First, write 
$\ket{\psi_{t-1}} = \left(\kets{\varphi_{t-1}^{q_1=0}} + \kets{\varphi_{t-1}^{q_1=1}}\right)\otimes\ket{t-1}_c$
where $\kets{\varphi_{t-1}^{q_1=s}}$ is the part of the work register 
with the control qubit $q_1$ in the state $\ket{s}$.
Define two extra states between $\ket{\psi_{t-1}}$ and $\ket{\psi_{t}}$:
\begin{eqnarray}
	\ket{\psi_{t}^1} &=& \kets{\varphi_{t-1}^{q_1=0}} \otimes \ket{l_1}_c 
										+ \phantom{X_{q_2}}\kets{\varphi_{t-1}^{q_1=1}} \otimes \ket{u_1}_c, \label{augment1}\\
	\ket{\psi_{t}^2} &=& \kets{\varphi_{t-1}^{q_1=0}} \otimes \ket{l_2}_c 
										+ X_{q_2} \kets{\varphi_{t-1}^{q_1=1}} \otimes \ket{u_2}_c. \nonumber
\end{eqnarray}
These states are again connected as a ``line'', because
$\bra{\psi_{t-1}}H_{\mathrm{sw}}^{(t)}\ket{\psi_{t}^1} = 
\bra{\psi_{t}^1}H_{\mathrm{sw}}^{(t)}\ket{\psi_{t}^2} =
\bra{\psi_{t}^2}H_{\mathrm{sw}}^{(t)}\ket{\psi_{t}}$ $= 1$.
We now use one railroad switch for every CNOT gate
\begin{eqnarray}
	H_{\mathrm{sw}} &=& 
			\sum_{t: \textrm{1-qubit }U_t} H_{F}^{(t)}
			+ \sum_{\textrm{CNOTs}} H_{\mathrm{sw}}^{(t)},
		\label{h3s}
\end{eqnarray}
and augment the subspace $\cH_{\mathrm{cmp}}^{\varphi_0}$ by the two extra states \eqref{augment1} for each of the gadgets. This results in a 3-local Hamiltonian \eqref{h3s}, whose restriction to the computational subspace is again \eqref{Hline}, generating a quantum walk on the augmented line \eqref{goodbasis}.
Note that for the single qubit gates, the operator $U_{t}\otimes \ket{t}\bra{t-1}_c$ is already 3-local
using the original pulse clock encoding.


\section{A qubit-qutrit switch}

\begin{figure}
	\begin{center}
	\includegraphics[width=2.8in]{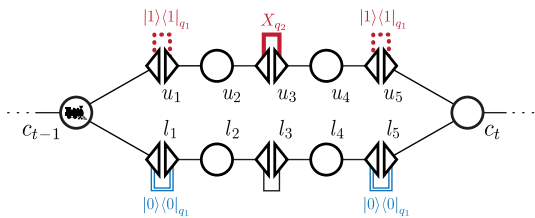} 
	\end{center}
	\caption{The 2-local {\em railroad switch} gadget made from qubits (circles) and qutrits (pairs of triangles).
	The transitions within a qutrit can be controlled by a train master, e.g. the internal transition $\ket{u_{1\mathtt{A}}}_c\leftrightarrow\ket{u_{1\mathtt{B}}}_c$ is allowed only when $q_1$ is $\ket{1}$. Only on the upper track, we flip of the target work qubit $q_2$ during the internal qutrit transition $\ket{u_{3\mathtt{A}}}_c\leftrightarrow\ket{u_{3\mathtt{B}}}_c$.
	\label{figure23}}
\end{figure}
If we allow the use of qutrits (e.g. spin-1 particles) in our system, we can decrease the locality of interactions 
from 3- 
to 2-local. The geometry of the qubit-qutrit interactions is depicted in Figure \ref{figure23}. 
We can view the two states of the clock qubits as an empty station $\ket{0}$, and a train $\ket{1}$ in it.
On the other hand, a qutrit station is either empty $\ket{\mathtt{O}}$, or has the train in one of its two stops $\ket{\mathtt{A}}$ or $\ket{\mathtt{B}}$. This allows a 2-local interaction to change the state of a work qubit while changing stations within a qutrit. In the 3-local railroad switch, the train master work qubit $q_1$ decided whether the train could pass to the upper/lower track (see Figure \ref{figureswitch}). Here, the controlled transitions happen between the internal stops of the qutrits $u_1, u_5, l_1, l_5$. In addition, we flip the target work qubit $q_2$ during the internal transition $\ket{u_{3\mathtt{A}}}_{c} \leftrightarrow \ket{u_{3\mathtt{B}}}_{c}$. 
The Hamiltonian for the upper track (see Figure \ref{figure23}) is
\begin{eqnarray}
	H_{23u}^{(t)} &=& \ket{u_{1\mathtt{A}}}\bra{t-1}_c 
									+ \ket{u_{1\mathtt{B}}}\bra{u_{1\mathtt{A}}}_c \otimes \ket{1}\bra{1}_{q_1} \label{toptrack}\\
									&+& \ket{u_2}\bra{u_{1\mathtt{B}}}_c 
									+ \ket{u_{3\mathtt{A}}}\bra{u_2}_c  
									+ \ket{u_{3\mathtt{B}}}\bra{u_{3\mathtt{A}}}_c \otimes X_{q_2} \nonumber\\
									&+& \ket{u_4}\bra{u_{3\mathtt{B}}}_c  
									+ \ket{u_{5\mathtt{A}}}\bra{u_4}_c  
									+ \ket{u_{5\mathtt{B}}}\bra{u_{5\mathtt{A}}}_c \otimes \ket{1}\bra{1}_{q_1} \nonumber\\
									&+& \ket{t}\bra{u_{5\mathtt{B}}}_c  
									+ c.c., \nonumber
\end{eqnarray}
where $\{\ket{u_{1\mathtt{A}}}_c, \ket{u_{1\mathtt{B}}}_c, \dots\}$ are the clock register states corresponding to the 8 possible positions of the train on the upper track. When the clock register has a single train in it, we can identify the clock states 1-locally (e.g. $\ket{u_{1\mathtt{B}}}\bra{u_{1\mathtt{B}}}_c = \ket{\mathtt{B}}\bra{\mathtt{B}}_{u_1}$).
The transition operators (e.g. $\ket{u_{3\mathtt{A}}}\bra{u_2}_c = \ket{0\mathtt{A}}\bra{1\mathtt{O}}_{u_2,u_3}$) can thus be implemented 2-locally. Moreover,
the operators involving a work qubit (e.g. $\ket{u_{5\mathtt{B}}} \bra{u_{5\mathtt{A}}}_c \otimes\ket{1}\bra{1}_{q_1}= \ket{\mathtt{B}}\bra{\mathtt{A}}_{u_5}\otimes\ket{1}\bra{1}_{q_1}$)
can also be made 2-local. Therefore, all the terms in $H_{23t}^{(t)}$ are 2-local, involving at most one qubit and one qutrit. The Hamiltonian $H_{23l}^{(t)}$ for the lower track is analogous to \eqref{toptrack}, with corresponding
lower-track clock states $\{\ket{l_{1\mathtt{A}}}_c, \dots\}$ and a simple term $\ket{l_{3\mathtt{B}}}\bra{l_{3\mathtt{A}}}_c$ (without flipping $q_2$).
Now replace each 3-local switch $H_{\mathrm{sw}}^{(t)}$ \eqref{Hswitch} with
\begin{eqnarray}
	H_{23}^{(t)} = H_{23u}^{(t)}+H_{23l}^{(t)}. \label{H23}
\end{eqnarray}
The basis of the computational subspace $\cH_{\mathrm{cmp}}^{\varphi_0}$ again needs to be augmented, similarly to what we did in \eqref{augment1}. After $\ket{\psi_{t-1}}$, we write the 8 states 
\begin{eqnarray}
	\ket{\psi_{t}^{r}} &=& \kets{\varphi_{t-1}^{q_1=0}} \otimes \ket{l_r}_c 
										+ \phantom{X_{q_2}} \kets{\varphi_{t-1}^{q_1=1}} \otimes \ket{u_r}_c, \label{augment23}\\
	\ket{\psi_{t}^{s}} &=& \kets{\varphi_{t-1}^{q_1=0}} \otimes \ket{l_s}_c 
										+ X_{q_2} \kets{\varphi_{t-1}^{q_1=1}} \otimes \ket{u_s}_c, \nonumber
\end{eqnarray}
with $r\in \{1\mathtt{A},1\mathtt{B},2,3\mathtt{A}\}$ and $s\in \{3\mathtt{B},4,5\mathtt{A},5\mathtt{B}\}$.
Moreover, we have the two ``blind-alley'' states with the train trying to use the track where it doesn't belong
\begin{eqnarray}
	\ket{\psi_{t}^{1x}} &=& 
										\kets{\varphi_{t-1}^{q_1=0}} \otimes \ket{u_{1\mathtt{A}}}_c 
										+ \phantom{X_{q_2}} \kets{\varphi_{t-1}^{q_1=1}} \otimes \ket{l_{1\mathtt{A}}}_c,
										 \label{blind}\\
	\ket{\psi_{t}^{5x}} &=& 
											\kets{\varphi_{t-1}^{q_1=0}} \otimes \ket{u_{5\mathtt{A}}}_c 
										+ X_{q_2} \kets{\varphi_{t-1}^{q_1=1}} \otimes \ket{l_{5\mathtt{A}}}_c. \nonumber
\end{eqnarray}
\begin{figure}
	\begin{center}
	\includegraphics[width=3in]{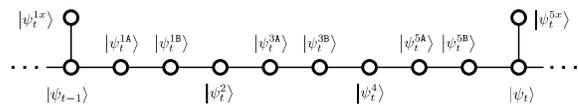} 
	\end{center}
	\caption{The geometry of the connections of the computational basis states \eqref{augment23}-\eqref{blind} implied by the transitions in the Hamiltonian \eqref{H23} for the qubit-qutrit 2-local railroad switch.
	\label{figurecomb}}
\end{figure}
The Hamiltonian \eqref{H23} connects these states with transitions whose geometry is depicted in Figure \ref{figurecomb}.
It is a line as we have seen before, with the two blind-alley states \eqref{blind}. The quantum walk dynamics induced on this graph is again similar to the quantum walk on a line, with mixing time (in the time-averaged sense) on the order of $O(L^2)$ for a convergence parameter choice $\ep=L^{-1}$, up to logarithmic factors) \cite{NecklaceWalks}.  

The crucial fact about this construction is the restriction of the computation to the computational subspace. 
Problematic states coming from the wrong initial state, states with more than one (or simply none) trains 
and bound states on the insides of the ``wrong'' tracks are not contained in $\cH_{\mathrm{cmp}}^{\ket{\varphi_0}}$,
making our life much easier.


\section{A qutrit from two qubits}
Finally, we take the qubit-qutrit 2-local gadgets and construct the 2-local qubit-qubit Hamiltonian
using an entangled encoding of clock states similar to the one in \cite{New3local}. We 
map the states of the qutrit into states of two qubits as
\begin{eqnarray}
	\ket{\mathtt{O}} \, & \rightarrow & \, \ket{00}, \nonumber\\
	\ket{\mathtt{A}} \, & \rightarrow & \, \ket{+} = \frac{1}{\sqrt{2}}\left(\ket{01}+\ket{10}\right),\\
	\ket{\mathtt{B}} \, & \rightarrow & \, \ket{-} = \frac{1}{\sqrt{2}}\left(\ket{01}-\ket{10}\right). \nonumber
\end{eqnarray}
Instead of the 2-local qutrit-qubit operators of the form $\left(\ket{\mathtt{B}}\bra{\mathtt{A}} + \ket{\mathtt{A}}\bra{\mathtt{B}}\right)\otimes V_d$,
we now use 
\begin{eqnarray}
	\frac{1}{2}\left(Z_1 - Z_2\right) \otimes V_d,
\end{eqnarray}
made of two 2-local qubit-qubit terms. Here $Z_1$ and $Z_2$ act on the two clock qubits encoding the qutrit,
and $V_d$ acts on a work qubit.
This operator annihilates states of the form $\ket{00}\ket{\varphi}$ and $\ket{11}\ket{\varphi}$, 
while inducing a transition
$
	\ket{+} \ket{\varphi} \,\leftrightarrow \, \ket{-}\left( V_d \ket{\varphi}\right)
$
using two-qubit entangled clock states $\ket{+}$ and $\ket{-}$.
In the Hamiltonian $H_{23}$, the active qutrit states also appear in the transitions such as
$\ket{1}_{c_{t-1}}\ket{\mathtt{O}}_{u_1} \leftrightarrow \ket{0}_{c_{t-1}}\ket{\mathtt{A}}_{u_1}$. 
Here, this particular transition becomes
$\ket{1}_{c_{t-1}}\ket{00}_{u_1,u'_1} \leftrightarrow \ket{0}_{c_{t-1}}\ket{+}_{u_1,u'_1}$,
and is implemented by
\begin{eqnarray}
	H^{100}_{0+} &=& 
					\frac{1}{\sqrt{2}}
					\ket{0}\bra{1}_{c_{t-1}} \otimes
					\left( 
							\ket{1}\bra{0}_{u_1}+\ket{1}\bra{0}_{u'_1} \right) \\
					&+& \frac{1}{\sqrt{2}}
					\ket{1}\bra{0}_{c_{t-1}} \otimes
					\left( 
							\ket{0}\bra{1}_{u_1}+\ket{0}\bra{1}_{u'_1} \right), \nonumber
\end{eqnarray}
a Hamiltonian built from 2-local terms. 
Note that the states $\ket{0}\ket{00}$ and $\ket{0}\ket{-}$ are annihilated by it.
Similarly, we write a Hamiltonian $H^{001}_{-0}$ wherever the transition $\ket{00}\ket{1}\leftrightarrow\ket{-}\ket{0}$ is called for in the clock register.
Restricting ourselves to the computational subspace with a single train, 
this new qubit-qubit 2-local Hamiltonian works just as the qubit-qutrit Hamiltonian \eqref{H23}.


\begin{figure}
	\begin{center}
	\includegraphics[width=0.9in]{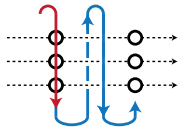} 
	\end{center}
	\caption{The geometrically local layout of our system. Each column represents $n$ work qubits, the full line depicts the clock register. As the clock progresses, first the gates are applied, and then the column of work qubits is swapped with the next one using CNOT gates, pushing the data to the right. Each qubit is involved in a constant number of 2-local interactions. The required number of work qubits is $nL$.
	\label{figurelayout}}
\end{figure}

The last necessary ingredient in the construction are single-qubit unitaries $W = \ket{w_0}\bra{w_0} +
e^{i \theta_w} \ket{w_1}\bra{w_1}$ which are not self-adjoint.
Let the target qubit be the train master, allowing the $\ket{w_1}$ branch on the upper track
and $\ket{w_0}$ on the lower track.
To make things simpler, we now use a simple pulse-clock encoding of the qutrit $u_3$
into two qubits $u_{3\mathtt{A}}$ and $u_{3\mathtt{B}}$ on the upper track (and similarly, on the lower track).
In the middle of the upper track, we then write a 2-local interaction term
\begin{eqnarray}
	H_{w_1} = \left(e^{i\theta_w} \ket{01}\bra{10} + 
	e^{-i\theta_w} \ket{10}\bra{01}\right)_{u_{3\mathtt{A}},u_{3\mathtt{B}}}
\end{eqnarray}
for the two clock qubits $u_{3\mathtt{A}}$ and $u_{3\mathtt{B}}$.
On the lower track, we use 
$\left(\ket{01}\bra{10} + \ket{10}\bra{01}\right)_{l_{3\mathtt{A}},l_{3\mathtt{B}}}$
without the phase shift.
The split into two tracks thus allows us to add a relative phase between them
which applies the single-qubit gate $W$ to the control work qubit as the computational paths rejoin.
The underlying unitary evolution in the computational subspace then again remains equivalent to the one induced by \eqref{H23}.

Finally, we can get geometric locality for this construction. The solution is to use $nL$ work qubits instead of only $n$, as in Figure \ref{figurelayout}, and wrapping the clock register around them four times per $n$ work qubits, reminiscent of \cite{OliveiraTerhal}. The winding (in 3D) is there to implement a round of gates, and then to perform a swap a work qubit column with a column of ancillae using 2 CNOT gates. The data thus moves to the next column, and the process continues. Each work qubit interacts with at most 5 switches, in some of them as a control and in some as a target. Altogether, it is involved in at most 28 two-local qubit-qubit interactions. This is far away from practical, but nevertheless a constant number of interactions per particle.


\section{Conclusions}

We presented a universal 2-local qubit HQC construction, which runs in three steps. A simple product state initialization, running for a time $O(L^2)$ (up to log factors) and a final computational basis measurement checking whether the clock register is in a state $\ket{t>L}_c$. The state of an output work qubit then contains the result of the quantum circuit $U$ acting on the work register of our initial state.
This construction differs from previous ones in two ways. First, the computation 
can run fast (does not rely on adiabatic evolution)
and second, the Hamiltonian is built from $O(nL)$ constant-norm terms
(does not use high-norm perturbation gadgets).

	
\section{Acknowledgments}
		This work was initiated at the 2009 ESI workshop on 
		Quantum Computation and Quantum Spin Systems. 
		DN thanks A.~Landahl, S.~Jordan, A.~Lutomirski, D.~Gosset and S.~Lloyd for fruitful discussions and gratefully acknowledges support from the project meta-QUTE ITMS 26240120022, the Slovak Research and Development Agency under the contract APVV LPP-0430-09,
		and the European Project Q-ESSENCE.
		


\end{document}